# Nanophotonic waveguide chip-to-world beam scanning at 68M Spots/s·mm²

Matt Saha[*,1], Y. Henry Wen,[*,1,2,α] Andrew S. Greenspon[*,1,2], Matthew Zimmermann[*,1], Kevin J. Palm[1,2], Alex Witte[1], Yin Min Goh[2], Chao Li[2], Jonathan Bumstead[3], Kevin Schädler[4], Mark Dong[1,2], Andrew J. Leenheer[6], Genevieve Clark[1,2], Gerald Gilbert[5, ⁑], Matt Eichenfield[6,7,&], and Dirk Englund[2,4†]

*These authors contributed equally to this work.
[1]The MITRE Corporation, 202 Burlington Road, Bedford, Massachusetts 01730, USA
[2]Research Laboratory of Electronics, Massachusetts Institute of Technology, Cambridge, Massachusetts 02139, USA
[3]The MITRE Corporation, 7525 Colshire Drive McLean, Virginia 22102, USA
[4]Axiomatic_AI, 501 Massachusetts Ave, Cambridge, MA 02139, USA
[5]The MITRE Corporation, 200 Forrestal Road, Princeton, New Jersey 08540, USA
[6]Sandia National Laboratories, P.O. Box 5800 Albuquerque, New Mexico 87185, USA
[7]College of Optical Sciences, University of Arizona, Tucson, Arizona 85719, USA
[α]hwen@mitre.org, [⁑]ggilbert@mitre.org, [&]eichenfield@arizona.edu, [†]englund@mit.edu

## Summary

A seamless chip-to-world photonic interface enables wide-ranging advancements in optical ranging, display, communication, computation, imaging, and light-matter interaction. An optimal solution allows for 2D scanning of a diffraction-limited beam from anywhere on a photonic chip over a large number of beam-spots in free-space. Currently, devices with direct PIC integration rely on tiled apertures with poor mode qualities, large footprints, and complex control systems. Micro-mechanical beam scanners have good beam quality but lack direct PIC integration and are inertially-limited due to the use of bulk optical components or structures in which the optical aperture and actuator sizes are inextricably linked, resulting in trade-offs among resolution, speed, and footprint. Here, we overcome these limitations with the photonic "ski-jump": a nanoscale optical waveguide monolithically integrated atop a piezoelectrically actuated cantilever which passively curls ~90° out-of-plane in a footprint of <0.1 mm², emits submicron diffraction-limited optical modes, and exhibits kHz-rate mechanical resonances with quality factors exceeding 10,000. Fabricated on a 200 mm wafer in a volume CMOS foundry, this device enables scalable 2D beam scanning with footprint-adjusted spot-rates of 68.6 mega-spots/s-mm² at CMOS-level voltages, which is equivalent to a 1 megapixel display at 100 Hz from a 1.5 mm² footprint, and exceeds the performance of state-of-the-art MEMS mirrors by >50✕. Using the photonic ski-jump, we demonstrate full-color image projection, video projection, and the initialization and readout of single photons from silicon vacancy centers in diamond waveguides. Based on current performance, we identify pathways for achieving >1 giga-spots at kHz-rates in a ~1 cm² area to provide a seamless, scalable optical pipeline between integrated photonic processors and the free-space world.

# Introduction

The transmission of information, from the astronomically large to the atomically small, is predominantly photonic. While the majority of our digital data is transmitted within photonic waveguide systems, a far larger datastream is transmitted photonically within and between the objects in the free-space world. An efficient chip-to-world photonic interface – the ability to directly convert between the time-bin modes of an integrated electro-optic processor and the spatial modes of the free-space world – opens up exciting opportunities for free-space communications and ranging[1–4], additive manufacturing[5], near-eye displays[6,7], biomedical imaging[8-9], machine learning[10], and atom control for quantum information[11]. However, our current digital infrastructure struggles with the immense data streams coming from real-world, free-space domains where every resolvable pixel is a separate channel that must be received, understood, and acted upon[12]. A similar challenge exists for quantum computing in the need for photonic control and readout of the millions of physical qubits that are required to reach universal fault tolerance[13]. Concurrently, photonic integrated circuits (PIC) have proliferated widely[14] and demonstrated increasingly sophisticated functionalities, including complex light conditioning for the control of atomic arrays[11] and free-space displays, in-physics algorithms[10], and deep co-learning at the edge[15]. As digital systems become more intelligent and agentic, the chip-to-world optical interface becomes a crucial link in the digital intelligence value chain.

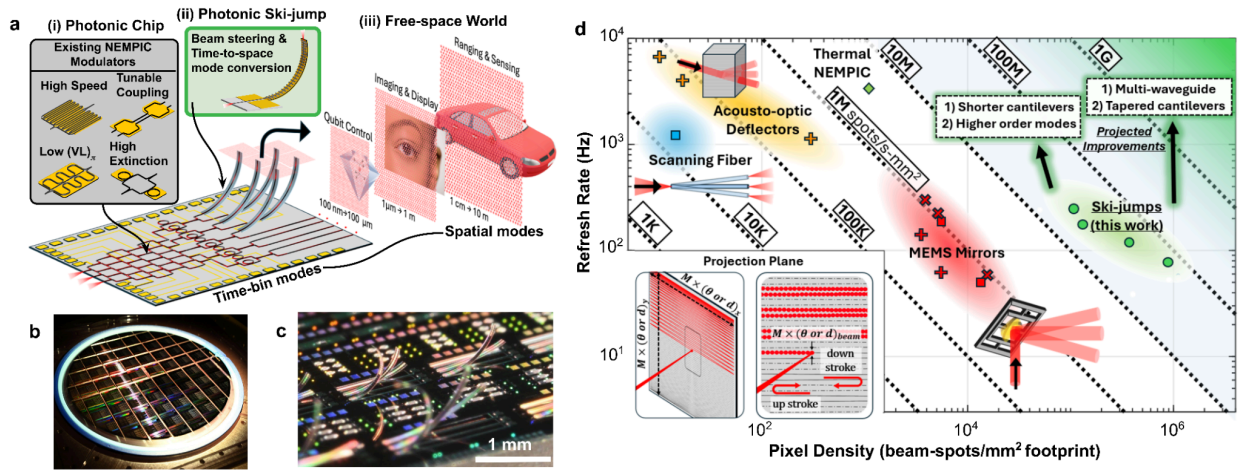

**Fig. 1 | Ski-jump enables fast time-to-space mode conversion as a scalable chip-to-word photonic interface, and its comparison to current beam scanning technologies. a,** (i) Existing piezoelectric nanoelectromechanical photonic integrated circuit (NEMPIC) components [11,16–19] enable fast photonic control and information processing over many time-bin modes on scalable photonics platform. (ii) Photonic ski-jump enables beam scanning and time-to-space mode conversion directly from the surface of a photonic chip. (iii) Targets in the free-space world have a large number of spatial modes with slow temporal evolution. **b,** A diced, unreleased NEMPIC wafer including optical modulators, couplers, and ski-jump devices. **c,** Various ski-jump device designs on a NEMPIC. **d,** Comparison of the photonic ski-jump with leading laser beam scanners as a function of footprint adjusted pixel density and refresh rate. Footprint refers to the active beam-scanning device area. Data points (green circles) are obtained from a single ski-jump measured in vacuum at 1, 2, 5, and 10 volts peak-to-peak (left to right). Lower left inset: projection plane FoV is given by the scan angle for pupil plane scanners or scan distance for focal plane scanners scaled by magnification M. AODs data points are from references [20–22]. MEMS mirrors data points are the highest performing devices from Table 4 in reference [1]. Scanning fiber is from reference [23] and thermal NEMPIC is from reference [24]

Despite this, the lack of a mode-efficient interface between the guided-wave mode-set of PICs and the continuous mode-set of free-space domains has prevented the seamless and scalable use of PICs in these areas. Integrated waveguide systems have a large number of time-bin modes by virtue of >100 GHz-rate electro-optic and THz-rate all-optical interactions[25]. But, they have a limited number of spatial waveguides per chip ($10^2$-$10^3$/mm) with broadband, diffraction-limited input/output (I/O) available only at the chip edge[26]. In contrast, the free-space domain has a nearly unbounded number of spatial modes[27] (~$10^{12}$/m$^2$) with relatively slow temporal variations (<$10^3$-$10^6$ Hz) for many relevant applications[1,6,28,29]. Despite the total mode-count being similar (~$10^{15}$-$10^{18}$), existing solutions fail to bridge this mode-set mismatch due to the lack of a diffraction-limited mode quality[30], wide field of view (FoV), fast scan rate, or direct, scalable coupling with PICs[1,6,21,23,29,31]. The ideal solution requires the ability to project and scan a diffraction-limited, single-mode beam to: *i*) a large number of resolvable beam-spots $N$ in the far field, *ii*) with a high refresh rate, *iii*) from a limited footprint, *iv*) directly on the surface of a programmable photonic chip.

Efforts in laser beam scanning have explored both tiled and continuous aperture designs, yet no single technology meets all performance criteria. Tiled aperture devices—spatial light modulators[27] (SLMs) and digital micromirror devices[31] (DMDs)—offer fully programmable beam control for dynamic phase and amplitude modulation over large apertures, enabling arbitrary beam patterns and rapid reconfiguration. However, their segmented aperture introduces diffraction that degrades far-field mode quality and spatial resolution[27,31]. In contrast, continuous aperture scanners—whether pupil-plane systems [like MEMS mirrors[1,32] and acousto-optic deflectors[20–22] (AODs)] or focal-plane scanners which directly emit guided modes from optical fibers or waveguides[8,23,33]—can project a single diffraction-limited beam-spot in the far field. But, these micro-mechanical scanners are inherently inertially limited which creates fundamental trade-offs among aperture size, device footprint, scan speed, and resolution. Notably, focal-plane scanners decouple numerical aperture from actuator size, offering a direct interface between guided and free-space modes[23,34]. For instance, scanning fibers are being commercialized for endoscopy[8] and near-eye displays[7,33] but still struggle with speed and resolution[35], and development so far has been limited to the use of bulk fibers and actuators[8,23].

Direct PIC-integrated beam scanning approaches—including optical phased arrays (OPAs) and hybrid MEMS-waveguide devices—offer compelling advantages. For example, Poulton et al.'s OPA[2,30] leverages mature silicon photonics to deliver rapid, fully programmable beam steering with wide FoVs and excellent electronic control, much like tiled aperture devices (e.g. SLMs and DMDs) that enable dynamic phase and amplitude modulation and arbitrary beam patterning. Azadeh et al.[24] made a key advance by integrating PIC waveguides atop thermal-MEMS scanners. However, both approaches share a common limitation: they rely on diffractive emitters that degrade far-field mode quality and spatial resolution while suffering from a limited operable wavelength range. While these methods underscore the strengths of direct PIC integration, their inherent diffraction-induced trade-offs highlight the need for novel architectures that enable scannable, non-diffractive emission across the surface of the chip.

Although each beam scanning technology offers its own set of unique advantages and trade-offs which can make direct comparison challenging, it is still instructive to distill performance into a single, reductive figure of merit (FoM) that captures the underlying beam scanning capacity. We quantify performance using two combined metrics: (1) the footprint-adjusted resolvable beam-spot count and (2) the refresh rate (see SI-8 for calculations). Their product yields a net FoM, expressed in spots/s-mm², which indicates the number of resolvable beam-spots that can be generated per second from a 1 mm² footprint of the active beam-scanning device area. This simplified metric provides a common baseline for comparing different technologies[20]. Conventional pupil-plane scanners require large apertures to minimize divergence and achieve high far-field resolution, forcing slow, high-power actuation that limits FoMs to ~500K–1M spots/s-mm². In contrast, focal plane scanners decouple optical and mechanical dimensions, but the use of bulk fibers and actuators has limited their FoM to < 50 K spots/s-mm². Thus, the scanning waveguide approach is hindered by the lack of scalable and scannable nanoscale, high-NA, single-mode waveguides directly integrated on a PIC.

Here, we introduce a new class of integrated photonic devices, the photonic "ski-jump", which overcomes these challenges and enables a scalable chip-to-world photonic interface.[36–40] This device, fabricated on a 200 mm wafer in a volume CMOS foundry, is composed of a nanoscale optical waveguide embedded monolithically on a piezoelectrically-actuated microcantilever with sub-microgram mass, ~2 μm thickness, and large out-of-plane curvature. The small mass and physical dimensions overcome the inertial limits of scanning fibers and break the FoM trade-offs of pupil plane scanners. The large upward curvature is achieved by engineering the directionality of the intrinsic material stress differential between the thin film layers of the cantilever bi-morph[41], an approach inspired by mechanical metamaterials[42] and which has been demonstrated on other quantum photonic platforms[43]. This provides vertical, scannable, broadband optical emission from anywhere on a 200 mm wafer with mechanical resonances from ~1 kHz to >100 kHz that significantly enhance the scan speed and FoV. The sub-micron integrated waveguides simultaneously reduce mass while increasing NA resulting in >1000× FoM improvement over existing fiber scanners[8,23] and >50× FoM improvement over mature MEMS mirror[1,29] and AODs[20,22].

Photonic ski-jumps are members of an unified family of piezo-actuated components on a CMOS-compatible nanoelectromechanical photonic integrated circuit (NEMPIC) platform. Prior works on this platform are shown in Fig. 1a and include tunable directional couplers[18], phase shifters[16,17], programmable MZIs[16,19], and tunable ring resonators[11,16]. This extensive process development kit (PDK) allows for complex photonic processing upstream of the ski-jump on the same monolithic photonic platform. Ski-jumps are also cryogenically-compatible for direct integration with solid-state qubit systems such as color centers in diamond which have been heterogeneously integrated onto the NEMPIC platform for microwave and strain control[44]. This opens up new routes for the addressing and readout of spin qubits. Future integration with electro-optic thin films[45] could enable 100 GHz modulation for the projection of sub-nanosecond optical pulses. The capabilities of the NEMPIC platform combined with the chip-to-world projection capability of the ski-jump enables scalable photonic and quantum control on- and off-chip.

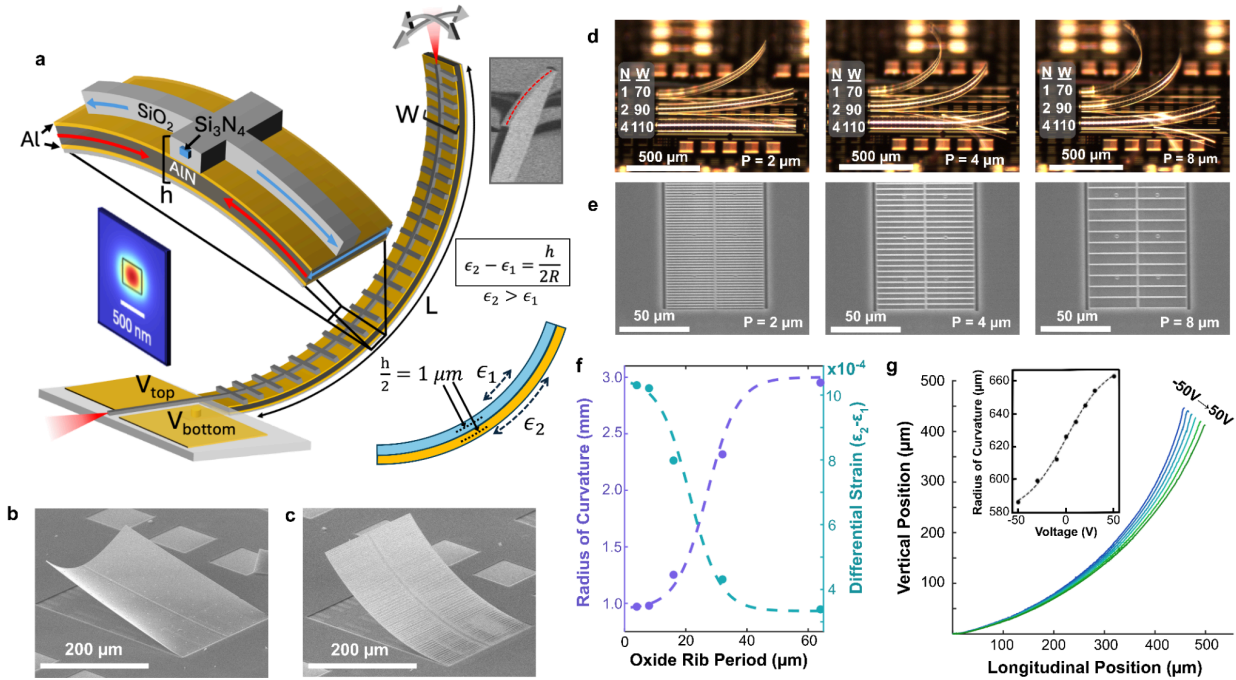

**Fig. 2 | Overview of the photonic ski-jump device. a,** Conceptual overview. Zoomed-in segment shows the cross-section composed of the lower layers ($SiO_2$-Al-AlN-Al) and upper optical layers ($Si_3N_4$-in-$SiO_2$). Each set of layers is ~1 μm thick resulting in a ~2 μm thick cantilever. Large curvature is obtained using cross-rib patterning of the top $SiO_2$ which expands laterally (blue arrows) causing downward lateral curvature and lateral compression (red) along with longitudinal expansion (blue) of the lower layers, resulting in upward longitudinal curvature. Upper right inset: scanning electron microscope (SEM) image demonstrating downward lateral curvature of the cantilever. Lower left inset: finite element method (FEM) simulation of the TE single-mode profile along the waveguide operating at 737 nm. Lower right inset: conceptual diagram of differential strain between the top and bottom layers. **b-c,** SEM images of cantilevers without (b) and with (c) cross-ribs to suppress lateral upward curvature and enhance longitudinal curvature. **d,** Variation of curvature with cross-rib period (P), cantilever width (W) in μm, and waveguide count (N). **e,** SEM images of cross-rib patterning with varying periods (P). **f,** Longitudinal curvature of cantilevers with 1 μm wide cross-ribs with periods ranging from 4 to 64 μm. **g,** DC actuation measured using white-light profilometry for -50 V to 50 V, (inset) radius of curvature for each dataset as a function of the applied DC voltage.

## Device overview

Crucial to the practical implementation of a chip-to-world photonic interface is scalability and mode quality. Photonic ski-jumps are fabricated using a 200 mm CMOS foundry process which combines silicon nitride ($Si_3N_4$) photonics with mature aluminum nitride (AlN) piezo-actuators as shown in Fig. 2a. Applying voltage across the Al electrodes induce piezo-electric stress in the AlN layer causing the ski-jump to deflect. The waveguides, composed of a $Si_3N_4$ core and silicon dioxide ($SiO_2$) cladding, are patterned at the top of the layer stack[17] and can be tailored for broadband, single-mode (or multi-mode) propagation across the visible-to-telecom spectrum[46]. The devices studied here have $Si_3N_4$ waveguides (400 nm wide and 300 nm thick) that are designed for single-mode propagation around 737 nm, the zero-phonon line of silicon vacancy (SiV) color center emitters in diamond. The spot-size, NA, and divergence of the ski-jump's optical output can be optimized for the intended application by tapering the waveguide width at the cantilever tip. For the devices discussed in this work, the waveguide width tapers down to 200 nm at the tip, resulting in a beam-spot size (i.e. mode field diameter) of $d_{spot,x}$=0.66 μm, $d_{spot,y}$=0.50 μm and divergence half-angles of $\theta_x$=41°, and $\theta_y$=53 for the fundamental transverse electric (TE) mode with further information provided in SI-9.

The cantilever curvature is achieved by engineering the gradient and directionality of the intrinsic stress of materials within the layer stack which expand upon release and bend the cantilever based on the bimorph mechanism[41] (Fig. 2a). To achieve upward curvature, the bottom layers must expand more than the top layers. This vertical stress gradient can be enhanced by patterning the top $SiO_2$ into periodic lateral rib structures ("cross-ribs") which has two key effects: 1) the large compressive stress of the waveguide is spread across the entire width of the cantilever while having minimal longitudinal expansion from the top $SiO_2$ layer, and 2) the bottom layers are laterally compressed which results in further longitudinal expansion and upward curvature. The strength of this mechanism is controlled by the cross-rib period and duty cycle (Fig. 2b-f) with an optimal period between 4 and 8 μm for a cross-rib width of 0.75 μm.

## DC and resonant 1D scanning of ski-jump cantilevers

Scanning a waveguide in space naturally implements conversion between waveguided time-bin modes and spatial modes in the continuum. DC actuation can be a useful means for tuning the ski-jump's curvature to optimize coupling to free-space optics, optical fibers, directly to free-space targets, or to other PICs. For DC voltages between -50 V and 50 V, a device with a length (L) of 800 μm and a width (W) of 90 μm has a vertical displacement range of 34.4 μm and a longitudinal displacement range of 29.4 μm (Fig. 2g). Each device also exhibits mechanical resonances that substantially enhance longitudinal and lateral displacement. The frequency and shape of these modes are determined by material stress, stiffness, and device geometry[47] with FEM results (Fig. 3a) that agree with the characteristic modes of a passively curled, singly clamped cantilever[47]. Most resonances within the simulated frequency range are longitudinal (Y) since the bending dimension is the film thickness. For lateral (X) modes, the bending dimension is the cantilever width which results in higher effective stiffness and resonance frequencies. Like other focal plane scanners, these devices scan both angle and displacement depending on the trajectory of the excited modes.

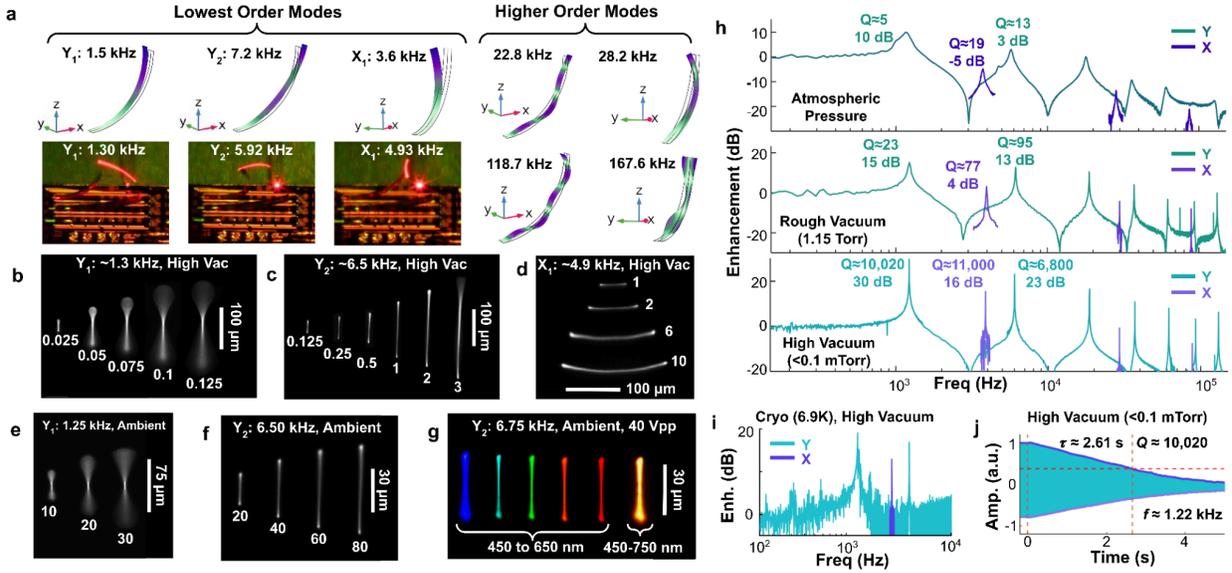

**Fig. 3 | Characterization of mechanical resonance modes. a,** FEM simulations for 7 of the resonant modes of a curled cantilever (L = 950 μm, W = 70 μm). Below: stroboscopic images of a resonantly driven ski-jump at its first two Y modes and first lateral mode. **b-f,** ICCD images of the waveguide output on resonance with various sinusoidal drive voltages (streak labels in units volts peak-to-peak [Vpp]) for a device with dimensions L=950 μm, W=70 μm. The exposure time is greater than the drive signal period so that the full range of motion is observed. Subpanels show the first two longitudinal modes and the first lateral mode in ambient and/or in high vacuum conditions. Approximate frequencies are given for (b-d) as resonance frequency shifts with voltage due to thermal redshifting. **g,** Broadband operation is demonstrated on a device with W=90 μm with a second-order longitudinal mode at 6.75 kHz. **h,** Small signal frequency response of Y and X beam displacement. Measurements were taken with the device at room temperature in atmospheric pressure, rough vacuum, high vacuum, and **i,** cryogenic conditions at high vacuum. Data is normalized to the Y displacement at low frequencies. **j,** Ring-down measurement for the fundamental longitudinal mode in high vacuum at room temperature.

We characterize the AC response of a device at various pressures and in cryogenic conditions (6.9 K) and driven using sinusoidal voltages with amplitudes up to 80 Vpp in ambient conditions and up to 10 Vpp in vacuum. The waveguide output is collected using an objective and imaged onto a CMOS camera or intensified charge coupled device (ICCD) for measurements of resonant displacement (Fig. 3b-g), or a position sensitive photodetector (PSD) for frequency response and ring-down measurements (Fig. 3h-j) with further detail on the experimental setup provided in SI-3.

On resonance, ski-jumps exhibit varying degrees of out-of-plane (Z) motion depending on the mechanical mode profile. The fundamental Y mode ($Y_1$) scans the cantilever tip along a nearly circular arc and exhibits a Y displacement exceeding 400 μm at 1 Vpp, as shown in stroboscopic imaging of the resonance mode profile (Supplementary Video 1) and in the ICCD images (Fig. 3b, 3e). In contrast, the higher order modes and, notably, the 2nd order Y mode ($Y_2$) exhibit much lower Z and angular motion due to the additional motional nodes. This separates the cantilever into distinct segments with opposing changes in curvature such that the composite motion results in a nearly flat tip trajectory while remaining nearly vertically oriented (Supplementary Videos 2 and Fig. 3c,f). We also find good agreement between the analytically derived expression for $FoM_{ID}$ and the data for the $Y_2$ mode (SI-8). Similarly, the fundamental X mode ($X_1$) exhibits a nearly flat tip trajectory due to the small passive curvature along the X direction (Supplementary Video 3 and Fig. 3d). Time-resolved motion of resonant beam scanning was also recorded using high-speed gating of the ICCD (Supplementary Video 4).

Decreasing the pressure enhances the resonant response (up to 30 dB) and reduces the linewidth. In high vacuum, the $Y_1$ mode has a quality ($Q$) factor of ≈10020 with displacements similar to ambient conditions ($Q≈5$) using >100✕ lower voltage. In vacuum, we observe a red-shifting nonlinearity due to electromechanical heating which passively stabilizes the drive frequency-resonance detuning, similar to other micro-oscillator systems[25,48]. A PID controller can provide additional long-term stability by feeding back on the phase of the device response. Furthermore, cryogenic operation is critical for applications in quantum information. At ~6.9 K in high vacuum, we observe a ~2✕ decrease in its radius of curvature due to differences in the rates of thermal expansion within the layer stack. Thus, ski-jumps intended for cryogenic applications can have half the footprint of their room-temperature counterparts.

The ski-jump's broadband transmission is demonstrated by scanning light from a supercontinuum source between 450 nm and 650 nm or white light (450 nm-750 nm), as shown in Fig. 3g. The propagation loss was measured to be 0.6-1.4 dB/mm depending on the cross-rib design (SI-11). These devices were designed with a cladding width of 1.4 µm that was further reduced to ~1.2 µm due to the $XeF_2$ release process which weakly etches $SiO_2$. Thus, the optical mode is overexposed to the cladding sidewalls which results in periodic scattering losses from the sharp index and strain gradient at the cross-ribs. Upon addressing this issue through widening of the waveguide cladding and rounding the cross-rib anchor points, we expect total propagation loss of the ski-jump to approach that of our standard routing waveguide: ~0.13 dB/mm, resulting in a total device loss of <1 dB.

We demonstrate the long term stability of our device at DC and resonance across multiple devices (SI-12). We show the device angle and position return after holding a voltage, and show the measured displacement is stable while driving devices for over 1 billion cycles and over dozens of hours. We note the amplitude of the $X_1$ mode drifts slightly over 10 hours, but this drift is slow and can be corrected by updating the driving frequency and amplitude.

## Resonant 2D beam scanning & image projection

At the core of a chip-to-world photonic interface is the ability to project a large 2D array of beam-spots in the far-field. To this end, we use a ski-jump with bilateral piezoelectric actuators (Fig. 4a). Each electrode can be driven at multiple frequencies with different voltages and relative phases to enable the excitation of both X and Y resonances simultaneously, resulting in Lissajous curves with varying refresh rates and beam-spot densities depending on the particular frequency ratio[28]. We characterize the X and Y frequency response of the device from 100 Hz to 50 kHz with separate, same-frequency signals sent to the 2 actuators while measuring the projected beam displacement with a PSD (Fig. 4b). Mechanically, driving out-of-phase signals cancels out the Y response while enhancing the X response. This enables cancellation of X-Y cross-coupling and efficient driving of the two orthogonal axes for 2D beam scanning. Optically, the combination of the 2nd order Y mode and fundamental X mode exhibits a sufficiently flat 2D scan area to serve as the basis for projecting a large 2D array diffraction-limited beam-spots. Though the curved trajectory of the cantilever results a spherical-type aberration across the scan area, correction of scanning fiber displays has been demonstrated using gradient-index optics and meta-lenses which provides a path for achieving an image plane that is flat to within a single rayleigh length of the beam-spot[7,8,49].

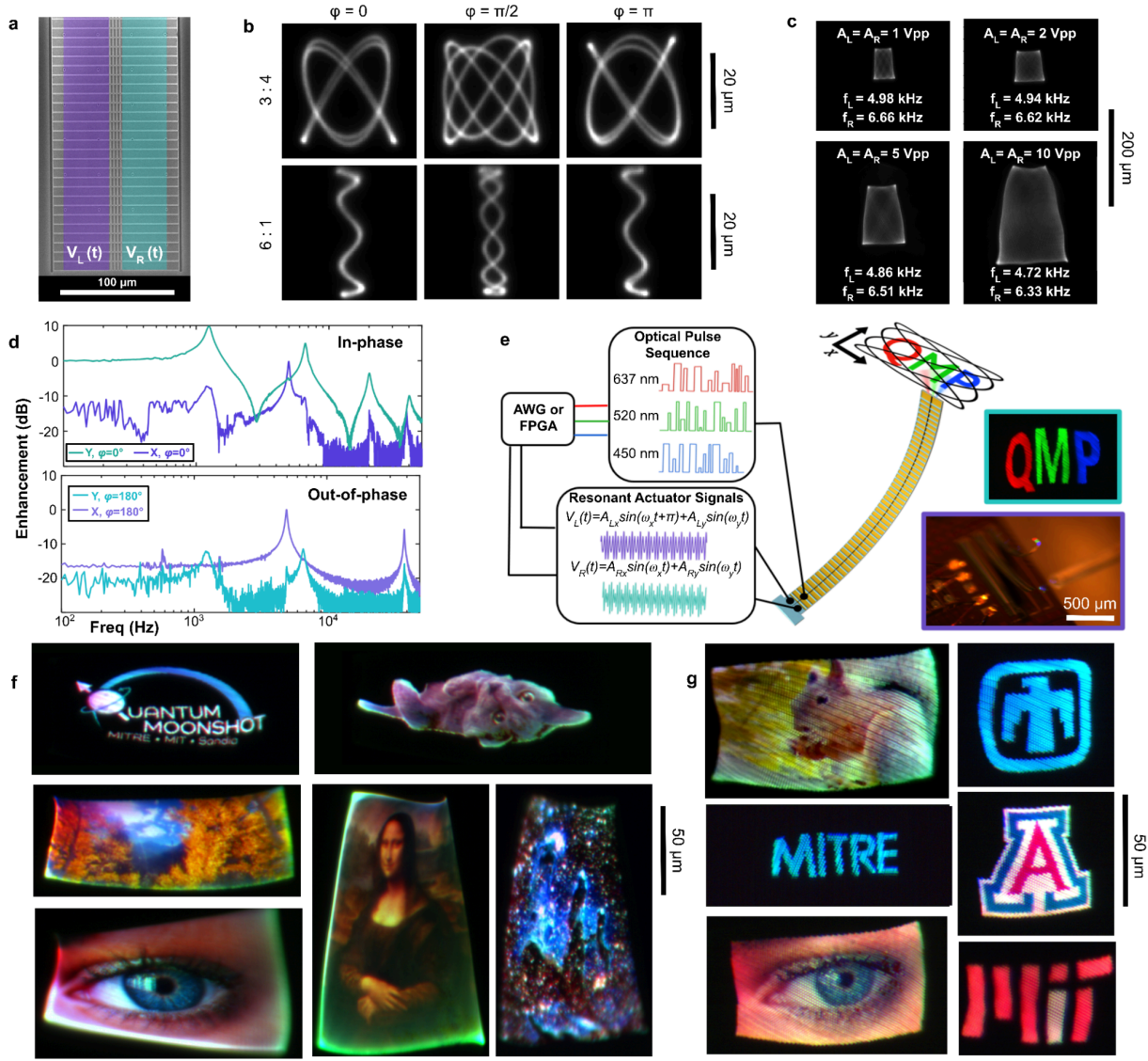

**Fig. 4 | Two-dimensional beam scanning with a split-electrode cantilever. a,** Top-down SEM image of a split-electrode cantilever's base. False color added to show left (purple) and right (cyan) electrodes. **b,** ICCD images of the split-electrode ski-jump's optical output tracing Lissajous patterns while the cantilever (L=950 μm, W=70 μm) is driven at different X:Y frequency ratios (4.83 kHz:6.44 kHz = 3:4 with a 1.61 kHz refresh rate, and 37.2 kHz:6.2 kHz = 6:1 with a 6.2 kHz refresh rate) and different phase offsets $\varphi$ between the X and Y drive signals. **c,** A demonstration of tuning the 2D scan's aspect ratio by varying amplitudes $A_L$ and $A_R$ with the left actuator driving near 4.83 kHz and the right actuator driving near 6.5 kHz in high vacuum conditions. Resonance peak frequencies shift depending on the voltage. **d,** Frequency response of the X and Y beam displacement while driving both actuators with in-phase and out-of-phase sinusoidal signals of 1 Vpp. Data is normalized to the in-phase Y displacement at low frequencies, and measurements were done in ambient conditions. **e,** Diagram of arbitrary 2D image projection with a split-electrode device. Purple inset: an image taken of the ski-jump projecting the letters "QMP" (Quantum Moonshot Project) in vacuum

conditions; 2 images, one of the stationary ski-jump and the other of the QMP projection with background lights off, are superimposed. Cyan inset: a top-down image of the QMP projection. **f,** Image projection using a long split-electrode device (L=1450 μm, W=70 μm) tracing high-fill lissajous patterns in vacuum conditions with $f_x$=1.561 kHz, $f_y$=2.639 kHz and a refresh rate of 7 Hz. **g,** Image projection examples using a shorter split-electrode device (L=950 μm, W=70 μm) in vacuum conditions with $f_x$=4.716 kHz, $f_y$=6.408 kHz, and a refresh rate of 36 Hz. 36 Hz video projections are provided as Supplementary Videos 6 and 7. Color projection examples were not modified post-capture. The camera's exposure time was set to match the ski-jump's refresh rate.

The choice of frequencies for the X and Y signals dictates both refresh rate and fill factor for the scan pattern. Applications such as LiDAR and image projection require high-fill, low-rate patterns, whereas low-fill, high-rate patterns are more suitable for optical control of atomic qubits. We generate high refresh rate Lissajous patterns by applying X and Y signals with low frequency ratios and adjusting their relative phase (Fig. 4c and Supplementary Video 5). Offsetting one of the frequencies sweeps the beam across the entire 2D area (Fig. 4d) with a low refresh rate which is equal to the greatest common factor of the X and Y frequencies. By pulsing the optical signal of the cantilevered waveguide with an off-chip modulator, we demonstrate the projection of full-color 2D images and video using these high-fill scans (Fig. 4e-f, Supplementary Videos 6 and 7) without the need for active stabilization, indicating that the trajectories are stable over time. We also demonstrate the long-term stability of both high-rate and high-fill patterns while driving for over 15 hours (SI-12), showing only slight drift that can be corrected by updating the amplitude and phase of the driving signal. While we currently modulate our optical signals upstream of our PIC, phase and amplitude modulation using on-chip modulators is readily achievable in future devices to enable all-on-chip projection systems for applications like LiDAR, AR/VR displays, phased array holographic projection, and qubit control. We also demonstrate compatibility with small form-factor commercial off-the-shelf optics by imaging the ski-jump using a iPhone 15 Pro main camera lens, achieving nearly 10× magnification over a 5‑cm optical track length and producing a uniform beam-spot of ~2 μm. Additionally, ZEMAX simulations indicate that with realistic design optimizations, molded multi-element lens stack of less than 5 mm can provide up to 30× magnification within an optical track length of less than 2 cm, confirming the feasibility of a compact, mass‑producible system (see SI-13 for further details).

## Ski-jump excitation of silicon vacancy color centers

To achieve a truly universal chip-to-world photonic interface, the projected light must control matter at the quantum mechanical level; that is, it must be spatially and spectrally coherent. This ability is central for the photonic control of atomic qubits and for applications across spectroscopy, fluorescence-based microscopy, and coherent ranging and sensing. It is also practically essential for achieving high spatial and temporal resolution for laser-based annealing, materials processing, and surgery. To demonstrate this capability, we use a single ski-jump to control a large number of solid-state atom-like qubits. Previous work used a controllable MZI mesh to route input laser light into four discrete channels for spatial and time-resolved control of quantum memories[50]. In contrast, the ski-jump has a continuous 2D range of outputs to control quantum memories as nodes on a grid. As such, the number of channels available scales based on the ski-jump's range of motion and optical mode size.

We demonstrate this capability on a wire bonded, fiber packaged device using 1D beam scanning from the ski-jump at ambient pressure and temperature. A diamond quantum microchiplet (QMC) consisting of 8 waveguides with implanted negatively charged SiVs is placed in a Montana 4K cryostat overhanging a bare silicon substrate (Fig. 5a). Laser light resonant with the SiV zero phonon line (ZPL) around 737 nm is routed from the end of the ski-jump in free-space to the tips of the individual diamond waveguides.

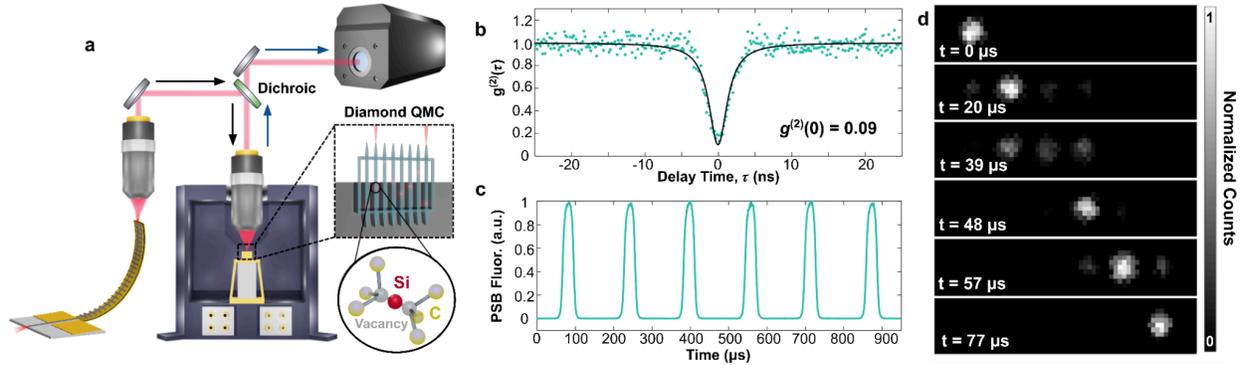

**Fig. 5 | Modulation of a single photon source using resonantly driven ski-jump devices. a,** Experimental setup for projecting the ski-jump's optical output onto a diamond QMC with implanted SiVs and readout onto a photodetector (APD or ICCD). **b,** Second-order autocorrelation measurement of a single emitter in one of the diamond waveguides. $g^{(2)}(0) = 0.09(9)$, indicating that a single emitter is being addressed. **c,** Time-dependent phonon sideband (PSB) fluorescence of the single emitter excited via the cantilevered waveguide oscillating at 6.34 kHz. The emitter is positioned at the apex of the device's range of motion. The extinction ratio is 27.5 dB with a repetition time of 157 μs. **d,** Ensemble fluorescence measurements collected with an ICCD. The driven cantilever addresses emitters in six distinct waveguides. The third waveguide, imaged at t = 39 μs, has a damaged taper. As a result, the resonant excitation was mostly scattered.

We align the emission of the static ski-jump to one of the diamond waveguide channels and tune the excitation laser to excite the C transition of a frequency-resolved single SiV. We verify that a single emitter is being addressed by performing a second-order autocorrelation measurement with $g^{(2)}(0) = 0.09(9)$, well below the 0.5 threshold for single photon emission (Fig. 6b). To periodically initialize this emitter, the ski-jump is driven to oscillate at 6.34 kHz while the SiV phonon sideband (PSB) fluorescence is routed to an avalanche photodiode coupled to a time-tagger (Fig. 5c). We observe consistent sideband emission with an extinction ratio of 27.5 dB, currently limited by scattered light. The standard deviation of the integrated pulse area is 0.003. We also demonstrate the control of multiple emitters in different waveguide channels by tuning the laser to a frequency that excites ensemble SiVs and scanning the ski-jump's optical output across the channels. We collect the PSB signal from each waveguide onto a high-speed ICCD, showing their real-time counts as the beam-spot scans over them (Fig. 5d).

This demonstration represents the missing link towards scalable, high-fidelity control of atomic qubits. Practical architectures will likely require arrays of smaller ski-jumps which can cyclically and simultaneously excite a subset of the qubit array with sub-$\mu$s refresh rates and dynamical shifting of the scan trajectory to interlink the qubit sub-arrays. Scaling towards such ski-jump arrays is considered in the discussion section. Generating optical π-pulses can be achieved by heterogeneous integration of electro-optic thin films[45] for >100 GHz bandwidth modulation with combined extinction ratio exceeding 50 dB. Additionally, emission from the diamond waveguides can couple to the underlying silicon nitride waveguides of the NEMPIC platform[44], for photon routing to ski-jumps for optical I/O. Since a ski-jump is fundamentally a 1xN switch, through the use of appropriate imaging optics, the combination of two ski-jump arrays enables out-coupling and routing of single photons for entanglement generation in a scheme akin to an NxM photonic matrix switch.

## Discussion

As demonstrated in this work, breaking the inertial trade-off between actuator mass and optical aperture opens access to beam-spot densities beyond conventional scanners. The one-dimensional scaling laws—$FoM_{ID,Y} \propto QL/(n^2Wh^2d_{spot,y})$ and $FoM_{ID,X} \propto QL/(n^2W^2hd_{spot,x})$—show that longer, thinner, and narrower cantilevers improve performance. A tapered cantilever with a wide base, narrow clamp, and light tip minimizes air damping akin to a whip while keeping $FoM_{ID,Y}$ constant and improving $FoM_{ID,x}$, corresponding to an increase in both resonant frequency and refresh rate. These trends motivate smaller, tapered ski-jumps capable of 1-2 order of magnitude improvements to the FoM. A complementary path routes multiple waveguides on a single cantilever, reducing the need for large lateral strokes and boosting the fill-factor × refresh-rate product. Whether achieved through improved actuator design, multiple waveguides, or both, the goal is a "self-covering" ski-jump that scans its own footprint with full-fill. A present array with ~25 % self-coverage yields ~25M spots/cm$^{-2}$, and full self-coverage would result in 100M spots/cm$^2$. A further 10× FoM improvement (≈1G spots/s-mm$^2$) would drive kHz-refresh "giga-spot" engines from a 10 cm² area.

However, collecting or projecting light from centimeter-scale ski-jump arrays poses challenges for standard optics. Fortunately, decades of high-volume manufacturing in precision-molded free-form optics have yielded near-diffraction-limited modulation transfer function (MTF) with sub-µm assembly tolerances at consumer costs. Modern phone camera optics exemplify these developments, and by leveraging optical reciprocity, the same micro-optics can collimate centimeter-scale arrays of ~1 µm ski-jump emitters across an equally wide field within a sub-cm³ module. Our measurements with the iPhone 15 Pro main lens validate this premise: a 5 cm track relay imaging setup produces 10× magnified image corresponding to ~2 µm spots, near the 1.67 µm resolving limit (S1-13). Rayleigh analysis (MTF=0.10) on Apple patent data[51] predicts ≈41M resolvable beam-spots across a 12.2 mm area, corresponding to >1000 ski-jumps.

However, these lenses are optimised for imaging to a flat sensor whereas ski-jumps sweep a shallow arc set by cantilever motion, risking defocus along the image periphery. Fortunately, the phone lens's moderate NA (≈ 0.3) provides a tens of microns depth of focus. Matching the ski-jump's optical output to this NA through further tapering of the ski-jump waveguide produces 1–2 µm spots-sizes. Though beam-spot density drops, the matched NA results in higher transmission and full-fill scanning can be achieved at faster speeds, thus preserving the FoM. We also explore optical compensation for the curved emission surface. Zemax simulations show that injection-molded free-form stacks tailored to this curvature can approach diffraction-limited performance. One 20 mm-long, 2 mm-diameter imaging design delivers ~30× magnification over an 18 mm track. A collimating variant of identical size achieves 0.13° divergence across 32° (0.18°/μm), while a 2 mm × 1 mm lens attains 0.23° over ±20° (±150 μm stroke, 0.13°/μm). In parallel, the curved trajectory can be addressed at the mechanical level by employing compound extension mechanisms[52], resonant mode shaping[47], or including waveguides terminated at staggered heights to functionally turn vertical motion into an intrinsic varifocal feature.

For monochromatic or narrow-band applications like eye-safe LiDAR, a single-element fish-eye metalens[53]—demonstrated at 940 nm and 1550 nm[54]—offers a compact, wafer-level route to near-diffraction-limited imaging over a sprawling 170° field. With <0.1° angular resolution (~2.4M addressable directions) across a >5 mm aperture, one metalens can service >175 present-day ski-jumps, reducing packaging to a single molded wafer or replicated polymer sheet. At the opposite scale, commercially available micro-lens arrays (MLAs)[55] provide lens-per-emitter control. When tiled above a ski-jump wafer, each MLA cell couples a single device—or a small cluster—to its own lenslet. This granular approach enables integral or light-field architectures in which every emitter–lens pair writes an elemental image, adding varifocal or volumetric capability to an otherwise planar array. An analysis of the projected beam-spot capacity and the optic volume of fully packaged systems is detailed in SI-14.

Additional practical constraints merit attention. Die-level vacuum packaging—already common in MEMS inertial sensors and available with optically clear glass lids—can provide the required low-pressure environment without bulky chambers[56,57]. Though current devices perform best under vacuum, the previously noted improvements in actuator design will narrow the gap with ambient pressure. Furthermore, resonant scanning, though energy-efficient, lacks true random access or static beam holding; yet, an array of >1000 emitters under the phone-lens footprint lets the neutral position be chosen simply by activating a different device, offering coarse, yet arbitrary, pointing capability. At higher optical powers, waveguide nonlinearities and thermal loading—familiar in other MEMS scanners[58]—can be mitigated with established control and heat-management techniques.

Despite these constraints, multiple engineering and optical paths remain open, underscoring the potentially transformative capabilities of the ski-jump platform. Achieving modest improvements in FoM—reaching 100M spots/s-mm²—would translate to the equivalent pixel count of twelve "4K" displays refreshed at 100 Hz within a 1 cm² area. For remote detection applications, employing the fish-eye meta-lens approach could enable approximately 1.83π steradian (170° field of view) coverage at 100 m with approximately 17.5 cm spatial resolution. Further application possibilities range from high-speed, high-resolution in vivo imaging to wide-field 3D printing and precision laser materials processing. The broad optical transparency window of silicon nitride, spanning from visible through telecom wavelengths, further enriches the applicability of ski-jump devices in diverse fields, from quantum optics and atomic control to environmental sensing and beyond.

## Conclusion

Ski‑jumps mark a significant advancement toward bridging the gap between integrated photonic circuits and the free‑space world. By directly integrating low-mass, high-NA waveguides onto a chip-based piezo-cantilever, we overcome key inertial and diffraction trade-offs that have long hindered conventional beam scanning technologies. Although our current implementations rely on resonant scanning in vacuum, these issues lay out clear avenues for future refinement. More importantly, establishing a robust chip-to-free-space photonic interface opens up exciting opportunities for machine vision and intelligence across applications such as autonomous vehicles, robotics, augmented reality, telepresence, and industrial automation. We anticipate this technology will form the basis of a light engine that enables machines to sense, communicate, and interact more effectively with their environment and with one another—ultimately enhancing human productivity and connectivity in our increasingly digital world.

## Methods

**Device fabrication and I/O**

The fabrication process follows the approach laid out in Stanfield et. al.[16]. A cross-section schematic is provided in Fig. S1. Cantilevers are fabricated on a CMOS platform with a silicon substrate. After an initial $SiO_2$ deposition, an amorphous silicon (a-Si) layer is deposited below the cantilever. The next layers are $SiO_2$, Al, AlN, Al, $SiO_2$, $Si_3N_4$, and lastly $SiO_2$. A defined etch around the cantilever exposes the underlying a-Si. Release holes are patterned every 30 μm along the length and width of the cantilever to ensure a complete release from the substrate. After wafer fabrication is complete, the PICs are placed in a xenon difluoride ($XeF_2$) gas etching chamber which etches away the a-Si, releasing the cantilever from the substrate except for a clamp at one end. The inherent stresses in the overlying thin films, along with tailored cross-rib patterning in the top $SiO_2$, causes the cantilever to curl out of the PIC plane. The standard $Si_3N_4$ waveguides used on these devices were 400 nm wide and 300 nm thick. There is an ~400 nm $SiO_2$ wide cladding buffer on the left and right sides of the waveguide(s), along with a top $SiO_2$ clad thickness of ~ 300 nm to 450 nm and a bottom $SiO_2$ clad thickness of ~850 nm.

Pads for electrical contact use routing metal and tungsten vias to independently route signals to the top and bottom electrodes of the single electrode cantilevers. The split-electrode devices share a ground plane but use independent signal pads to control the left and right actuators. Electrical signals are sent to these pads using GSG(SG) probes or via wirebonds to a custom PCB. Laser light is routed to the ski-jumps either using an edge-coupled lensed fiber or a fully packaged fiber-to-grating coupler system.

## Acknowledgements

Major funding for this work is provided by MITRE for the Quantum Moonshot Program. D.E. acknowledges partial support from Brookhaven National Laboratory, which is supported by the U.S. Department of Energy, Office of Basic Energy Sciences, under Contract No. DE-SC0012704 and the NSF RAISE TAQS program.

M.E. performed this work, in part, with funding from the Center for Integrated Nanotechnologies, an Office of Science User Facility operated for the U.S. Department of Energy Office of Science. The authors thank Ryan Fortin for his work on resonant control and stabilization using a PID controller. The authors also thank Linsen Li and Kevin Chen for their fabrication of the SiV-implanted diamond sample. The authors thank Nils Otterstrom, Sivan Trajtenberg-Mills, Thomas Propson, and Xiyuan Lu for valuable feedback on the manuscript. Y.M.G. and C.L. also acknowledge support from the U.S. Department of Energy, Office of Science, National Quantum Information Science Research Centers, Quantum Systems Accelerator. Y.H.W. acknowledges Juejun Hu and Zhaoyi Li for insightful discussion and provision of micro-optic components.

Sandia National Laboratories is a multimission laboratory managed and operated by National Technology & Engineering Solutions of Sandia, LLC, a wholly owned subsidiary of Honeywell International Inc., for the U.S. Department of Energy's National Nuclear Security Administration under contract DE-NA0003525. This paper describes objective technical results and analysis. Any subjective views or opinions that might be expressed in the paper do not necessarily represent the views of the U.S. Department of Energy or the United States Government.

## Author contributions

Y.H.W., M.D., and D.E. conceptualized the device; Y.H.W. performed the FoM analysis and state-of-the-art comparison with support from M.Z.; Y.H.W. performed the analytical mechanical beam theory and scaling analysis.; A.S.G. formalized the device design and created the device layout; A.S.G., Y.H.W., and M.S. performed device modeling and simulation; M.S. and A.S.G. designed and built the experimental setups with support from A.W. and Y.H.W.; M.S. performed device imaging and profilometry with support from Y.H.W. and A.S.G.; M.S. and M.Z. performed the frequency response measurements; M.S. performed further experimental mechanical mode analysis and DC characterization; M.Z. and M.S. performed 2D image and video projection with assistance from Y.H.W.; A.W. optically packaged the devices; K.J.P. performed the cryogenic color center excitation experiment with support from M.S. and A.S.G.; M.Z. ran the control electronics and wrote the control software; M.Z. performed ski-jump propagation loss measurements. Y.H.W. conducted micro-optics integration. C.L. and Y.M.G. contributed to device diagnosis, output mode profiling, and micro-optics integration. J.B. and K.S. created ZEMAX simulations for ski-jump imaging and micro-optics integration. M.Z.

performed ski-jump stability measurements. M.E. conceived the PIC platform and developed the fabrication architecture with A.J.L.; A.J.L. also supervised the fabrication process; G.C. assisted with sample preparation and developed post-fabrication tuning of the cantilever's passive curvature. M.E., G.G., and D.E. supervised all aspects of the project; G.G. supervises the Quantum Moonshot Program. Y.H.W., A.S.G., M.Z., M.S., and K.J.P. wrote the manuscript with input from all authors.

## Competing interests

D.E. is a Scientific Advisor to and is a co-founder of Axiomatic_AI. The other authors declare no competing interests.

## Data availability

The data that support the plots within this paper are available under restricted access due to MITRE's information security policies. Access can be obtained from the corresponding authors upon request.

## References

1. Wang, D., Watkins, C. & Xie, H. MEMS Mirrors for LiDAR: A review. *Micromachines (Basel)* **11**, (2020).

2. Poulton, C. V. *et al.* Coherent LiDAR With an 8,192-Element Optical Phased Array and Driving Laser. *IEEE J. Select. Topics Quantum Electron.* **28**, 1–8 (2022).

3. Zhang, X., Kwon, K., Henriksson, J., Luo, J. & Wu, M. C. A large-scale microelectromechanical-systems-based silicon photonics LiDAR. *Nature* **603**, 253–258 (2022).

4. Li, B., Lin, Q. & Li, M. Frequency-angular resolving LiDAR using chip-scale acousto-optic beam steering. *Nature* **620**, 316–322 (2023).

5. Corsetti, S., Notaros, M., Sneh, T., Page, Z. A. & Notaros, J. Silicon-photonics-enabled chip-based 3D printer. *Light: Science & Applications* **13**, 1–11 (2024).

6. Hsiang, E.-L. *et al.* AR/VR light engines: perspectives and challenges. *Adv. Opt. Photonics* **14**, 783 (2022).

7. Li, Z. *et al.* Meta-optics achieves RGB-achromatic focusing for virtual reality. *Sci Adv* **7**, (2021).

8. Lee, C. M., Engelbrecht, C. J., Soper, T. D., Helmchen, F. & Seibel, E. J. Scanning fiber endoscopy with highly flexible, 1 mm catheterscopes for wide-field, full-color imaging. *J. Biophotonics* **3**, 385–407 (2010).

9. Sacher, W. D. *et al.* Implantable photonic neural probes for light-sheet fluorescence brain imaging. *Neurophotonics* **8**, 025003 (2021).

10. Farmakidis, N., Dong, B. & Bhaskaran, H. Integrated photonic neuromorphic computing: opportunities and challenges. *Nat Rev Electr Eng* **1**, 358–373 (2024).

11. Menssen, A. J. *et al.* Scalable photonic integrated circuits for high-fidelity light control. *Optica* **10**, 1366–1372 (2023).

12. Kaufmann, E. *et al.* Champion-level drone racing using deep reinforcement learning. *Nature* **620**, 982–987 (2023).

13. Fowler, A. G., Mariantoni, M., Martinis, J. M. & Cleland, A. N. Surface codes: Towards practical large-scale quantum computation. *Phys. Rev. A* **86**, 032324 (2012).

14. Bogaerts, W. & Rahim, A. Programmable Photonics: An Opportunity for an Accessible Large-Volume PIC Ecosystem. *IEEE J. Sel. Top. Quantum Electron.* **26**, 1–17 (2020).

15. Sludds, A. *et al.* Delocalized photonic deep learning on the internet's edge. *Science* **378**, 270–276 (2022).

16. Stanfield, P. R., Leenheer, A. J., Michael, C. P., Sims, R. & Eichenfield, M. CMOS-compatible, piezo-optomechanically tunable photonics for visible wavelengths and cryogenic temperatures. *Opt. Express* **27**, 28588–28605 (2019).

17. Dong, M. *et al.* High-speed programmable photonic circuits in a cryogenically compatible, visible–near-infrared 200 mm CMOS architecture. *Nat. Photonics* **16**, 59–65 (2021).

18. Wen, Y. H. *et al.* Tunable Directional Couplers in a Scalable Piezo-MEMS Platform. in *Frontiers in Optics + Laser Science 2023 (FiO, LS)* (Optica Publishing Group, Washington, D.C., 2023). doi:10.1364/fio.2023.fth1e.4.

19. Dong, M. *et al.* Piezo-optomechanical cantilever modulators for VLSI visible photonics. *APL Photonics* **7**, 051304 (2022).

20. Römer, G. R. B. E. & Bechtold, P. Electro-optic and acousto-optic laser beam scanners. *Phys. Procedia* **56**, 29–39 (2014).

21. A review of physical principles and applications of acousto-optic deflectors on the basis paratellurite. *Physics & Astronomy International Journal* **3**, 62–65 (2019).

22. ISOMET XY 2D AOD. Preprint at https://isomet.com/PDF%20acousto-optics_deflectors/Data%20sheets%20-%20defbluered/ODXY1441-T100S-3.pdf.

23. Wang, W.-C. *et al.* Mirrorless MEMS imaging: a nonlinear vibrational approach utilizing aerosol-jetted PZT-actuated fiber MEMS scanner for microscale illumination. *Microsyst Nanoeng* **10**, 13 (2024).

24. Sharif Azadeh, S. *et al.* Microcantilever-integrated photonic circuits for broadband laser beam scanning. *Nat. Commun.* **14**, 2641 (2023).

25. Joshi, C. *et al.* Thermally controlled comb generation and soliton modelocking in microresonators. *Opt. Lett.* **41**, 2565 (2016).

26. Xiang, C., Jin, W. & Bowers, J. E. Silicon nitride passive and active photonic integrated circuits: trends and prospects. *Photonics Res.* **10**, A82 (2022).

27. Panuski, C. L. *et al.* A full degree-of-freedom photonic crystal spatial light modulator. *Nat. Photonics* **16**, 834–842 (2022).

28. Wang, J., Zhang, G. & You, Z. Improved sampling scheme for LiDAR in Lissajous scanning mode. *Microsyst Nanoeng* **8**, 64 (2022).

29. Hofmann, U., Janes, J. & Quenzer, H.-J. High-Q MEMS Resonators for Laser Beam Scanning Displays. *Micromachines* **3**, 509–528 (2012).

30. Hashemi, H. A review of semiconductor-based monolithic optical phased array architectures. *IEEE*

*Open J. Solid State Circuits Soc.* **1**, 222–234 (2021).

31. Zhang, B., Peng, P., Paul, A. & Thompson, J. D. A scaled local gate controller for optically addressed qubits. *Optica* (2023) doi:10.1364/OPTICA.512155.

32. Hofmann, U. *et al.* Resonant biaxial 7-mm MEMS mirror for omnidirectional scanning. *JM3.1* **13**, 011103 (2013).

33. Schowengerdt, B. T. & Watson, M. D. Ultra-high resolution scanning fiber display. *World Patent* (2014).

34. Khayatzadeh, R., Civitci, F., Ferhanoglu, O. & Urey, H. Scanning fiber microdisplay: design, implementation, and comparison to MEMS mirror-based scanning displays. *Opt. Express* **26**, 5576–5590 (2018).

35. Guttag, K. Magic Leap Fiber Scanning Display (FSD) – 'The Big Con' at the 'Core'. *KGOnTech* https://kguttag.com/2018/01/06/magic-leap-fiber-scanning-display-fsd-the-big-con-at-the-core/ (2018).

36. Wen, Y. H. *et al.* Photonic control of atom-like qubits using 2D scanning waveguide-on-cantilever 'ski-jumps'. in *Quantum 2.0 Conference and Exhibition* vol. 16 QTh2A.7 (Optica Publishing Group, Washington, D.C., 2024).

37. Saha, M. *et al.* High-speed off-chip beam steering via photonic integrated waveguides embedded on vertical ski-jump cantilevers. in *Frontiers in Optics + Laser Science 2023 (FiO, LS)* FTu6E.2 (Optica Publishing Group, Washington, D.C., 2023).

38. Zimmermann, M. *et al.* Resonant mode analysis and 2D projection via waveguide-on-cantilever ski jumps. in *Frontiers in Optics + Laser Science 2024 (FiO, LS)* FTh3C.3 (Optica Publishing Group, Washington, D.C., 2024).

39. Wen, Y. H. *et al.* Stress-programmable out-of-plane interposers for 3-D photonic integration & control. in *Frontiers in Optics + Laser Science 2024 (FiO, LS)* FM4B.5 (Optica Publishing Group, Washington, D.C., 2024).

40. Wen, Y. H. *et al.* Full-color chip-to-free-space nanophotonic scanning waveguide display at 68 Million spots/(s-mm^2). in *Optical Architectures for Displays and Sensing in Augmented, Virtual, and Mixed Reality (AR, VR, MR) VI* (eds. Hua, H., Argaman, N. & Nikolov, D. K.) 103 (SPIE, 2025).

41. Timoshenko, S. Analysis of bi-metal thermostats. *J. Opt. Soc. Am.* **11**, 233 (1925).

42. Chen, S., Chen, J., Zhang, X., Li, Z.-Y. & Li, J. Kirigami/origami: unfolding the new regime of advanced 3D microfabrication/nanofabrication with 'folding'. *Light: Science & Applications* **9**, 1–19 (2020).

43. Qvotrup, C. *et al.* Curved GaAs cantilever waveguides for the vertical coupling to photonic integrated circuits. *Opt. Express* **32**, 3723–3734 (2024).

44. Clark, G. *et al.* Nanoelectromechanical Control of Spin–Photon Interfaces in a Hybrid Quantum System on Chip. *Nano Lett.* **24**, 1316–1323 (2024).

45. Valdez, F., Mere, V. & Mookherjea, S. 100 GHz bandwidth, 1 volt integrated electro-optic Mach–Zehnder modulator at near-IR wavelengths. *Optica* **10**, 578 (2023).

46. Wen, Y. H. *et al.* Strain-concentration for fast, compact photonic modulation and non-volatile memory. *Optica* **11**, 1511 (2024).

47. Rao, S. S. *Vibration of Continuous Systems*. (John Wiley & Sons, Nashville, TN, 2007).

48. Carmon, T., Yang, L. & Vahala, K. Dynamical thermal behavior and thermal self-stability of microcavities. *Opt. Express* **12**, 4742–4750 (2004).

49. Majumdar, A. Large field-of-view polychromatic metalens for full-color scanning fiber endoscopy. *Research Square* (2024) doi:10.21203/rs.3.rs-4283076/v1.

50. Palm, K. J. *et al.* Modular chip-integrated photonic control of artificial atoms in diamond nanostructures. *Optica* (2023) doi:10.1364/OPTICA.486361.

51. Yao, M. S. Y. S. Near-Infrared Imaging Lens (Apple Inc). https://patentimages.storage.googleapis.com/e7/90/f6/e2f66f46b3e75e/US9869847.pdf.

52. Xie, X. & Livermore, C. A high-force, out-of-plane actuator with a MEMS-enabled microscissor

motion amplifier. *J. Phys. Conf. Ser.* **660**, 012026 (2015).

53. Shalaginov, M. Y. *et al.* Single-element diffraction-limited fisheye metalens. *Nano Lett.* **20**, 7429–7437 (2020).

54. 2Pi Optics Inc. (Private Communications, April 2025).

55. Ingeneric Micro-lens Array Catalog. Preprint at https://ingeneric.com/wp-content/uploads/2023/05/INGENERIC_MLA_2023.pdf.

56. Schott Hermetic Optical Enclosures Catalog. Preprint at https://mss-p-009-delivery.stylelabs.cloud/api/public/content/bd9907539d224c7a81c020ea6e06d70c?v=7a7343ed&download=true.

57. Edinger, P. *et al.* Vacuum-sealed silicon photonic MEMS tunable ring resonator with an independent control over coupling and phase. *Opt. Express* **31**, 6540–6551 (2023).

58. *MEMS Mirrors Technical Note*. https://www.hamamatsu.com/content/dam/hamamatsu-photonics/sites/documents/99_SALES_LIBRARY/ssd/mems_mirror_koth9003e.pdf (2023).